\documentstyle[aps,twocolumn,graphicx,epsfig]{revtex}

\begin{document}

\draft
\title{Expansion of a coherent array of Bose-Einstein condensates}
\author{P. Pedri, L. Pitaevskii$^*$, S. Stringari}
\address{Dipartimento di Fisica, Universit\`a di Trento and Istituto
Nazionale per la Fisica della Materia, I-38050 Povo, Italy}
\address{$^*$Kapitza Institute for Physical Problems, Moscow, Russia}
\author {C. Fort, S. Burger, F. S. Cataliotti, P. Maddaloni, F. Minardi,
M. Inguscio}
\address{LENS, Dipartimento di Fisica, Universit\`a di Firenze and Istituto
Nazionale per la Fisica della Materia, L.go E. Fermi 2, I-50125
Firenze, Italy}

\date{\today}
\maketitle

\begin{abstract}
\noindent We investigate the properties of a coherent array
containing about $200$ Bose-Einstein condensates produced in a
far detuned $1D$ optical lattice. The density profile of the gas,
imaged after releasing the trap, provides information about the
coherence of the ground-state wavefunction. The measured atomic
distribution is characterized by interference peaks. The time
evolution of the peaks, their relative population as well as the
radial size of the expanding cloud are in good agreement with the
predictions of theory. The $2D$ nature of the trapped condensates
and the conditions required to observe the effects of coherence
are also discussed.
\end{abstract}

 \narrowtext

Coherence is one of the most challenging features exhibited by
Bose-Einstein condensates. On the one hand it underlies the
superfluid phenomena exhibited by these cold atomic gases.  On
the other hand it characterizes in a unique way their matter wave
nature at a macroscopic level.  Coherence requires that the
system be characterized by a well defined phase, giving rise to
interference phenomena.

After the first interference measurements carried out on two
expanding condensates at MIT \cite{andrews97} the experimental
study of interference in Bose-Einstein condensed gases has become
an important activity of research opening the new field of
coherent atom-optics.

 The possibility of confining Bose-Einstein condensates in
optical lattices has opened further perspectives in the field
\cite{kasevich98}. Bose-Einstein condensates confined in an
optical standing-wave provide in fact a unique tool to test at a
fundamental level the quantum properties of systems in periodic
potentials. The observation of interference patterns produced by
an array of condensates trapped in an optical lattice was already
used as a probe of the phase properties of this
system~\cite{kasevich01,haensch01} also allowing to proof the
phase relation in an oscillating Josephson current~\cite{jose}.
In~\cite{kasevich01} the interference effect has been used to
explore the emergence of number squeezed configurations in
optically trapped condensates.

The main purpose of this paper is to investigate the ground state
properties of  the system of a fully coherent array of
condensates. To this aim we have explored the interference
pattern in the expanded cloud, reflecting the initial geometry of
the sample.

The basic phenomenon we want to explore is the atom optical
analog of light diffraction from a grating.  The analogy is best
understood considering a periodic and coherent array of identical
condensates aligned along the $x$-axis.  In momentum space the
order parameter takes the form

\begin{eqnarray}
\Psi(p_x)=\Psi_0(p_x) \sum_{k=0, \pm 1 .. \pm k_M}
e^{\frac{ikp_xd}{\hbar}} \nonumber \\
=\Psi_0(p_x) \frac{\sin[(2k_M+1)p_xd/2\hbar]}{\sin p_xd/2\hbar}
\label{momenti}
\end{eqnarray}
where $k$ labels the different sites of the lattice, $2k_M+1$ is
the total number of sites (in the following we will assume $k_M
\gg 1$) and $d$ is the distance between two consecutive
condensates.  The quantity $n_0(p_x) =|\Psi_0(p_x)|^2$ is the
momentum distribution of each condensate (see
Eq.~(\ref{Psi0sigma}) below).  The momentum distribution of the
whole system, given by $n(p_x)=\mid\Psi(p_x)\mid^2$, is affected
in a profound way by the lattice structure and exhibits
distinctive interference phenomena. Actually the effects of
coherence are even more dramatic than in the case of two
separated condensates \cite{PS}. Indeed, in the presence of the
lattice the momentum distribution is characterized by sharp peaks
at the values $p_x=n2\pi \hbar/d$ with $n$ integer (positive or
negative) whose weight is modulated by the function $n_0(p_x)$.
Furthermore, differently from the case of two separated
condensates, interference fringes appear only if the initial
configuration is coherent. In principle the momentum distribution
can be directly measured {\it in situ} using 2-photon Bragg
spectroscopy. This possibility has been already implemented
experimentally for a single condensate \cite{mitnp}. However, the
very peculiar structure of (\ref{momenti}) is expected to
influence in a deep way also the expansion of the atomic cloud
after the release of the trap. The width of the central peak
($n=0$) of the momentum distribution is of the order $\Delta p_x
\sim \hbar/R_x$ where $R_x \sim k_Md$ is half of the length of
the whole sample in the $x$-direction and the corresponding
atomic motion, after the release of the trap, will be
consequently slow.  On the other hand the peaks with $n\ne 0$
carry high momentum and the center of mass of these peaks will
expand fast according to the asymptotic law
\begin{equation}
x(t)=\pm n\frac{2 \pi \hbar}{d m} t \; .
\label{xt}
\end{equation}
The occurrence of these peaks is the analog of multiple order
interference fringes in light diffraction.

We create an array of BECs of $^{87}$Rb in the $|F=1, m_F=-1
\rangle$ state by superimposing the periodic optical potential
$V_{opt}$ of a far detuned standing-wave on the harmonic
potential $V_B$ of the magnetic trap. For a more detailed
description see \cite{EPL,PRLsuperfluid}.  The resulting
potential is given by
\begin{eqnarray}
&&V=V_B+V_{opt} \nonumber \\ &&= \frac{1}{2} m
\left(\omega_x^2x^2+ \omega_{\perp}^2(y^2+z^2)\right) + sE_R
\cos^2(q x+{\pi \over 2}) \label{optpot}
\end{eqnarray}
with $m$ the atomic mass, $\omega_x=2 \pi \times 9$~Hz and
$\omega_{\perp}=2 \pi \times 92$~Hz the axial and radial
frequency of the magnetic harmonic potential and $x$ lying in the
horizontal plane. In (\ref{optpot}) $s$ is a dimensionless factor,
$q=2 \pi/ \lambda$ is the wavevector of the laser light creating
the standing wave and producing local minima in $V_{opt}$
separated by $d =\lambda/2$ and $E_R=\hbar^2 q^2/2m \sim 2\pi
\hbar \times 3.6$~kHz is the recoil energy of an atom absorbing
one lattice photon. By varying the intensity of the laser beam
(detuned 150~GHz to the blue of the D$_1$ transition at
$\lambda=795$~nm) up to 14~mW/mm$^2$ we can vary the intensity
factor $s$ from 0 to 5. We calibrated the optical potential
measuring the Rabi frequency ($\Omega_R$) of the Bragg transition
between the momentum states $-\hbar q$ and $+\hbar q$ induced by
the standing wave. The intensity factor is then given by $s=2
\hbar \Omega_R/E_R$ \cite{salomon}.

The procedure to load the condensate in the combined
(magnetic+optical) trap is the following: we load $^{87}$Rb atoms
in the magnetic trap and cool the sample via rf-forced
evaporation until a significant fraction of condensed atoms is
produced. We then switch on the laser standing-wave and continue
the  evaporative cooling to a lower temperature (T$ \ll$T$_c$).
Typically, the BEC splits over $\sim 200$ wells, each containing
$100\sim 500$ atoms. After switching off the combined potential
we let the system expand and take an absorption image of the
cloud at different expansion times $t_{exp}$.

\begin{figure}
\begin{center}
\includegraphics[width=8cm]{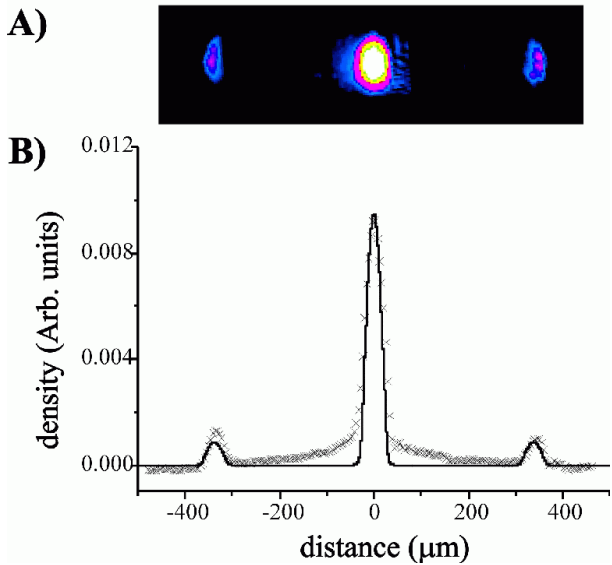}
\caption{A) Absorption image of the density distribution of the
expanded array of condensates. B)  Experimental density profile
(crosses) obtained from the absorption image (A) integrated along
the vertical direction. The wings of the central peak result from
a small thermal component. The continuous line corresponds to the
calculated density profile for the expanded array of condensates
for the experimental parameters ($s=5$ and $t_{exp}=29.5$~ms).}
\label{foto}
\end{center}
\end{figure}

\begin{figure}
\begin{center}
\includegraphics[width=8cm]{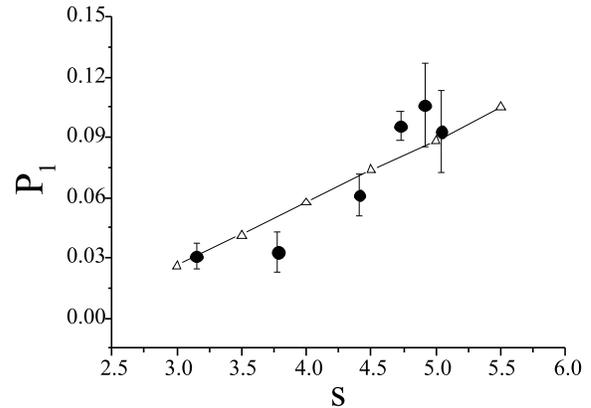}
\caption{Experimental (circles) and theoretical (triangles)
values of the relative population of the $n=1$ peak with respect
to the $n=0$ central one as a function of the intensity factor
$s$ of the optical potential $V_{opt}$.} \label{efficienza}
\end{center}
\end{figure}

In Fig.~\ref{foto}A we show a typical image of the cloud taken at
t$_{exp}=29.5$~ms, corresponding to a total number of atoms $N=
20000$ and to a laser intensity  $s=5$. From the images taken
after the expansion we can determine the relative population of
the lateral peak with respect to the central one. The
experimental results for the relative population of the first
lateral peak as a function of the laser intensity $s$ are shown
in Fig.~\ref{efficienza}.

The structure of the observed density profiles is well reproduced
by the free expansion of the ideal gas where the time evolution
of the order parameter, in coordinate space, takes the form:

\begin{equation}
\Psi(x,t)=\frac{1}{(2 \pi)^3} \int dp_x \Psi(p_x) e^{i p_x
x/\hbar} e^{-ip_x^2t/2m\hbar} \; .
\end{equation}

For a realistic description of $\Psi(p_x)$ we have improved the
simple ansatz (\ref{momenti}) in order to account for the
$k$-dependence of the number of atoms $N_k$ contained in each
well.  Due to magnetic trapping, the central condensates with $k
\ll k_M$ will be in fact more populated than the ones occupying
the sites at the periphery.  We have accounted for the modulation
by the simple law $N_k = N_0 (1 - k^2/k^2_M)^2$ which will be
derived below. For $\Psi_0$ we have made the Gaussian choice

\begin{equation}
\Psi_0(p_x) \propto \exp[-p^2_x\sigma^2/2\hbar^2]
\label{Psi0sigma}
\end{equation}
corresponding, in coordinate space, to $\Psi_0(x) \sim
\exp[-x^2/2\sigma^2]$.   Using (\ref{Psi0sigma}) it is immediate
to find that the relative population of the $n \neq 0$ peaks with
respect to the central one ($n=0$) obeys the simple law
\begin{equation}
P_n=\exp[-4 \pi^2 n^2 \sigma^2/d^2]
\label{pop}
\end{equation}
holding also in the presence of a smooth modulation of the atomic
occupation number $N_k$ in each well.  Result (\ref{pop}) shows
that, if $\sigma$ is much smaller than $d$ the intensity of the
lateral peaks will be high, with a consequent important layered
structure in the density distribution of the expanding cloud. The
value of $\sigma$, which characterizes the width of the
condensates in each well, is determined, in first approximation,
by the optical confinement. The simplest estimate is obtained by
the harmonic expansion of the optical potential (\ref{optpot})
around its minima: $V=\sum_k (1/2)m\tilde{\omega}_x^2(x-kd)^2$
with $\tilde{\omega}_x=2\sqrt{s}E_R/\hbar$, yielding $\sigma =
d/(\pi s^{1/4})$. However this estimate is not accurate except
for very intense laser fields. A better value is obtained by
numerical minimization of the energy using the potential
(\ref{optpot}) and the wavefunction (\ref{Psi0sigma}). This gives
$\sigma/d = 0.30 $, $ 0.27 $, and $0.25$ for $s= 3$, $4$ and $5$
respectively. The predicted results for the density distribution
$n(x)$ $=|\Psi(x) |^2$ evaluated for $s=5$ and $t=29.5$~ms are
shown in Fig.~\ref{foto}B (continuous line).

From the above calculation we can also determine  the relative
population $P_{n}$ of the $n=1$ peak as a function of the
intensity factor $s$. This is shown in Fig.~\ref{efficienza}
together with the experimental results. The good comparison
between experiment and theory reveals that the main features of
the observed interference patterns are well described by this
simple model.

The $1D$ model discussed above can be generalized to $3D$ through
the ansatz
 \begin{equation}
\Psi_0({\bf r}) = \sum_{k=0, \pm 1 .. \pm k_M}
e^{-(x-kd)^2/2\sigma^2} \Psi_k({\bf r}_{\perp}) \label{psi0G}
\end{equation}
which can be used, through a variational calculation, to describe
the ground state of the system in the presence of the optical
potential, magnetic trapping and two-body interactions.  For
sufficiently intense optical fields the value of $\sigma$ is not
significantly affected by two-body interactions, nor by magnetic
trapping. On the other hand interactions are important to fix the
shape of the condensate wave function in the radial direction.
Neglecting the small overlap between condensates occupying
different sites and using the Thomas-Fermi approximation to
determine the wave function in the radial direction we obtain the
result
 \begin{equation}
 \mid \Psi_k({\bf r}_{\perp})\mid^2 ={\sqrt{2} \over g}
 \mu_k\left(1- {r^2_{\perp} \over (R_{\perp})^2_k}\right)
\label{Psik}
 \end{equation}
where $(R_{\perp})_k = \sqrt{2\mu_k/m\omega^2_{\perp}}$ is the
radial size of the $k$-th condensate, $g$ depends on the
scattering length $a$ through the relation $g=4 \pi \hbar^2 a /m$,
while
\begin{equation}
\mu_k =  {1\over 2} m \omega_x^2d^2 \left(k^2_M-k^2\right)
\label{muk}
\end{equation}
plays the role of an effective $k$-dependent chemical potential.
The value of $k_M$ is fixed by the normalization condition $N=
\sum N_k$ and is given by
\begin{equation}
k^2_M = {2 \hbar \overline{\omega} \over m\omega^2_xd^2} \left({15
\over 8 \sqrt\pi}N{a\over a_{ho}}{d\over \sigma}\right)^{2/5} \; .
\label{kM}
\end{equation}
In (\ref{kM})
$\overline{\omega}=(\omega_x\omega_{\perp}^2)^{1/3}$ is the
geometrical average of the magnetic frequencies, $a_{ho} =
\sqrt{\hbar/m\overline{\omega}}$ is the corresponding oscillator
length and $a$ is the $s$-wave scattering length.  From the above
equations one also obtains the result $N_k = N_0(1-k^2/k^2_M)^2$
with $N_0 = (15/16)N/k_M$. Equations (\ref{Psik}-\ref{kM})
generalize the well known Thomas-Fermi results holding for
magnetically trapped condensates \cite{RMP} to include the
effects of the optical lattice.

Neglecting two-body interaction terms in the determination of the
Gaussian width in the $x$-direction is a good approximation only
if $\mu_k$ is significantly smaller than the energy $\hbar
\tilde{\omega}_x$.  This condition is rather well satisfied in the
configurations of higher lattice potential employed in the
experiment.  For example, using the typical parameter $N=5 \times
10^4$ for the total number of atoms and the values
$\overline{\omega} = 2\pi \times 42$~Hz and $a/a_{ho} = 3.2
\times 10^{-3}$, we find, for $s=4$, $ \tilde{\omega}_x \sim 2\pi
\times 14$~kHz, $\mu_{k=0} \sim 2\pi\hbar \times 0.5 $~kHz and
$k_M \sim 100$, corresponding to $N_0 \sim 500$. Notice that with
these values the condition $\mu_k \gg \hbar \omega_{\perp}$
required to apply the Thomas-Fermi approximation is rather well
satisfied for the central wells.

The fact that $\mu_k$ turns out to be significantly smaller than
$\hbar \tilde{\omega}_x$ not only explains why the interference
patterns emerging during the expansion are well described by the
ideal $1D$ model for the array used above, but also points out the
$2D$ nature of the condensates confined in each well. In this
context it is worth pointing out that the bidimensionality of
these condensates is ensured up to temperatures of the order of
$k_BT \sim \hbar \tilde{\omega}_x$, which is significantly higher
than the expected value of the critical temperature for
Bose-Einstein condensation. Our sample can then be used also to
explore the consequence of the array geometry on the critical
phenomena exhibited by these optically trapped Bose gases
\cite{burger}.

\begin{figure}
\begin{center}
\includegraphics[width=8cm]{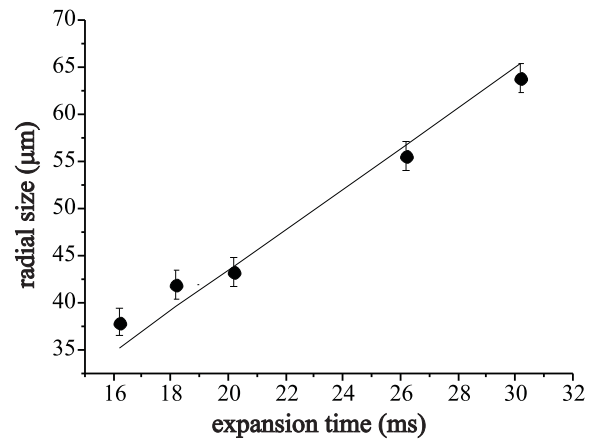}
\caption{Radial size of the central peak as a function of the
expansion time. Experimental data point are compared with the
expected asymptotic law $R_{\perp}=R_{\perp} (0) \omega_{\perp}
t_{exp}$.} \label{raggi}
\end{center}
\end{figure}

The above discussion permits also to  explain the behaviour of
the radial expansion of the gas. In the presence of the density
oscillations produced by the optical lattice the problem  is not
trivial and should be solved numerically by integrating the GP
equation. However, after the lateral peaks are formed, the
density of the central peak expand smoothly according to the
asymptotic law $R_{\perp}(t) = R_{\perp}(0)\omega_{\perp}t_{exp}$,
holding for a cigar configuration in the absence of the optical
lattice \cite{castin}. In Fig.~\ref{raggi} the linear law is
plotted using  the expression $R_{\perp}(0) \sim (R_{\perp})_{k=0}
= k_M d \omega_x/\omega_{\perp}$ derivable from Eq.~(\ref{muk})
for the condensate occupying the central well. This choice for
$R_{\perp}(0)$ is justified if  the population of the lateral
interference peaks is  small so  that their creation does not
affect the radial expansion of the system.

Let us finally discuss the conditions required for our system to
exhibit coherence. At zero temperature the coherence between two
consecutive condensates in the array is ensured if $E_c \ll E_J$,
where $E_c$ and $E_J$ are the parameters of the Josephson
Hamiltonian for two adjacent condensates \cite{leggett}. In
particular $E_c = 2
\partial \mu_k /\partial N_k$ is the interaction parameter while $E_J
= (\hbar^2/m) \int d{\bf
r}_{\perp}[\Psi_k\partial\Psi_{k+1}/\partial x-
\Psi_{k+1}\partial\Psi_k/\partial x]_{x=0}$ is the Josephson
parameter describing the tunneling rate through the barrier
separating two consecutive wells. In our case ($s=4$ and
$N=5\times 10^4$) we find $E_c \sim 2\pi \hbar \times 1$~Hz for
the most relevant central condensates ($k \ll k_M$) while, by
solving numerically the Schr\"odinger equation in the presence of
the optical potential $V_{opt}$ of Eq.~(\ref{optpot}), we find
$E_J \sim 2\pi \hbar \times 600 $~kHz. The value of $E_J$ is so
large that one can safely conclude that the ground state of the
system is fully coherent and that the effects of the quantum
fluctuations of the phase will be consequently negligible. This
reflects the fact that, even for the largest values employed for
the laser power the overlap between consecutive condensates is
not small enough. The value of $E_J$ is also much higher than the
values of $k_BT$ used in the experiment, so that also the effects
of the thermal fluctuations of the phase of the condensate can be
ignored.  This suggests that the fringes associated with the
expansion of the condensate will remain visible up to the highest
values of $T$, corresponding to the critical temperature for
Bose-Einstein condensation.  We have carried out experiments at
different values of $T$ where the signal obtained by imaging the
expanding cloud can be naturally decomposed in two parts: an
incoherent component due to the thermal cloud which is
parametrized by a classical Gaussian Boltzmann distribution, and
a Bose-Einstein component exhibiting the interference effects
discussed above. In our experiment the interference peaks are
visible up to $k_BT \sim 2\pi \hbar \times 4.2$~kHz. In order to
point out the effects of the fluctuations of the phase one should
lower the value of $E_J$ by orders of magnitude. This can be
achieved by increasing significantly the laser power generating
the optical lattice. Such effects have been recently observed in
the experiment of \cite{kasevich01}.

In conclusion, we have investigated the consequences of coherence
on the properties of an array of Bose-Einstein condensates. We
have observed peculiar interference patterns in the density of the
expanded cloud, reflecting the new geometry of the sample and
discussed on a theoretical basis some key features exhibited by
these optically trapped gases. Further studies in this direction
include the possible effects of thermal decoherence
\cite{pitaevskii01,tognetti} in the presence of tighter optical
traps and the emergence of $2D$ effects in the thermodynamic
properties of these novel systems.

This work has been supported by the EU under Contracts No. HPRI-CT
1999-00111 and No. HPRN-CT-2000-00125, by the MURST through the
PRIN 1999 and PRIN 2000 Initiatives and by the INFM Progetto di
Ricerca Avanzata ``Photon matter''.

\begin{thebibliography}{99}
\bibitem{andrews97} M.~R.~Andrews {\it et al.}, Science
{\bf 275}, 637 (1997).
\bibitem{kasevich98} B.~P.~Anderson and M.~A.~Kasevich, Science {\bf 282},
1686 (1998).
\bibitem{kasevich01} C.~Orzel {\it et al.}, Science {\bf 291}, 2386
(2001).
\bibitem{haensch01} M.~Greiner {\it et al.}, cond-mat/0105105.
\bibitem{jose} F.~S.~Cataliotti {\it et al.}, Science accepted
for publication.
\bibitem{PS} L. Pitaevskii and S. Stringari, Phys. Rev. Lett. {\bf 83}, 4237 (1999).
\bibitem{mitnp} D.~M.~Stamper-Kurn {\it et al.}, Phys. Rev. Lett. {\bf 83},
2876 (1999).
\bibitem{EPL} C.~Fort {\it et al.}, Europhys. Lett. {\bf 49}, 8
(2000).
\bibitem{PRLsuperfluid} S.~Burger {\it et al.}, Phys. Rev. Lett. {\bf 86}, 4447
(2001).
\bibitem{salomon} E.~Peik et al., Phys. Rev. A {\bf 55}, 2989
(1997).
\bibitem{RMP} F.~Dalfovo, S.~Giorgini, L.~P.~Pitaevskii, and S.~Stringari,
Rev. Mod. Phys. {\bf 71}, 463 (1999).
\bibitem{burger} S.~Burger {\it et al.} in preparation.
\bibitem{castin} Y.~Castin and R.~Dum, Phys. Rev. Lett. {\bf 77}, 5315
(1996).
\bibitem{leggett} A.~J.~Leggett, Rev. Mod. Phys. {\bf 73}, 307
(2001).
\bibitem{pitaevskii01} L.~Pitaevskii and S.~Stringari,
cond-mat/0104458.
\bibitem{tognetti} A.~Cuccoli, A.~Fubini, V.~Tognetti, and
R.~Vaia, cond-mat/0107387
\end{thebibliography}
\end{document}